\documentclass[conference,letterpaper]{IEEEtran}
\IEEEoverridecommandlockouts
\usepackage{balance}

\usepackage{mathtools}
\allowdisplaybreaks[4] % 0..4, higher = more willing to break

\usepackage{cuted}
\usepackage{amsmath,amssymb,amsfonts,latexsym}
\usepackage{afterpage}
\usepackage{stfloats}
\usepackage{relsize} % For your \mathsmaller command

\usepackage{algpseudocode}

\usepackage{textcomp}

\usepackage{diagbox}
\usepackage{comment}
\usepackage{enumitem}
\usepackage{multicol, blindtext}
\usepackage{lipsum}
\usepackage{bbm}
\usepackage[linesnumbered,vlined,ruled]{algorithm2e}
\usepackage{xcolor, soul}
\sethlcolor{yellow}
\usepackage{multirow}
\usepackage{bbding}
\usepackage{graphicx}
\usepackage{xfrac}

\usepackage{tabularx}

\def\BibTeX{{\rm B\kern-.05em{\sc i\kern-.025em b}\kern-.08em
    T\kern-.1667em\lower.7ex\hbox{E}\kern-.125emX}}

\usepackage[skip=0pt]{caption}
\usepackage[font=footnotesize,skip=0pt]{subcaption}	
\begin{document}

\bstctlcite{IEEEexample:BSTcontrol}
\title{Latency-Aware Digital Twin-Assisted Cooperative Perception for Autonomous Vehicles

}
\author{
\IEEEauthorblockN{
Boniface~Uwizeyimana\IEEEauthorrefmark{1},
Manobendu~Sarker\IEEEauthorrefmark{2}, and 
Abraham~O.~Fapojuwo\IEEEauthorrefmark{1}
}

\IEEEauthorblockA{\IEEEauthorrefmark{1}Department of Electrical and Software Engineering, University of Calgary, Canada}

\IEEEauthorblockA{\IEEEauthorrefmark{2}Poly-Grames Research Center, Department of Electrical Engineering, Polytechnique Montr\'{e}al, Canada}

\IEEEauthorblockA{
\IEEEauthorrefmark{1}\{boniface.uwizeyimana, fapojuwo\}@ucalgary.ca, \IEEEauthorrefmark{2}manobendu.sarker@polymtl.ca
}
}

\maketitle

\IEEEpubidadjcol

\begin{abstract}
This paper introduces a digital-twin (DT)-assisted cooperative perception framework designed to improve perception accuracy under end‑to‑end (E2E) latency constraints and to balance perception accuracy and E2E latency under communication resource constraints in autonomous vehicles. We formulate an optimization problem that maximizes perception accuracy subject to latency and communication limitations, and solve it using a newly proposed coarse-to-fine search (CTFS) algorithm. Simulation results show that the proposed CTFS algorithm achieves 96.6\% perception accuracy, close to exhaustive search, under latency constraints while reducing computational complexity by approximately 85.78\%. The DT-assisted framework further achieves a 50\% reduction in the non-DT communication cost through estimated, time-synchronized state updates.
\end{abstract}


\section{Introduction}
Recent advancements in autonomous vehicles have brought cooperative perception to the forefront, as it enhances sensing capabilities by mitigating occlusions, improving robustness under adverse conditions, and extending the perception range \cite{ngo2023cooperative}. Autonomous vehicles rely on onboard sensors such as light detection and ranging (LiDAR) and cameras to perceive their surroundings and support safe navigation. However, individual vehicles face inherent limitations due to restricted sensing ranges and environmental obstructions. Cooperative perception addresses these limitations by enabling vehicles to share sensory information, which is then fused by an aggregating agent to improve detection performance. Edge-assisted cooperative perception \cite{bai2024survey} further strengthens this capability by combining local observations with information shared by neighbouring vehicles, thereby enabling a more comprehensive environmental understanding. 
Achieving high perception accuracy is essential, but it must also satisfy strict real-time constraints to support safe and reliable operation. In this context, latency plays a critical role in determining the practical effectiveness of cooperative perception systems.

Existing approaches to cooperative perception generally operate at three levels of fusion: early, intermediate, and late fusion \cite{wei2025cooperative, urbaningv2x2025}. Early fusion involves transmitting raw sensor data to an aggregating agent, achieving high detection accuracy \cite{urbaningv2x2025}, but facing significant communication overhead and sensitivity to latency and noise,  which limits its practical deployment \cite{ngo2023cooperative}. Intermediate fusion allows vehicles to extract local features and send only feature maps, balancing accuracy and communication costs while offering improved robustness to latency. Late fusion shares detection results instead of raw data, reducing communication needs but typically resulting in lower accuracy due to limited use of sensory information \cite{urbaningv2x2025}.

Despite these developments, the problem of maximizing perception accuracy under strict latency constraints remains insufficiently explored. Prior work in \cite{hu2022where2comm} introduced a communication-aware strategy that selects the spatial area of interest for transmission using a communication confidence threshold, thereby reducing communication overhead. This approach is further evaluated in \cite{urbaningv2x2025} using the UrbanIng dataset. Nevertheless, determining the optimal transmission strategy in real time remains challenging when latency constraints and resource limitations must be jointly satisfied. The key challenge lies in selecting the communication confidence threshold and allocating transmission and computational resources to enhance perception accuracy while ensuring timely decision-making. However, such decisions are often driven by instantaneous system observations, which limits their ability to anticipate latency violations caused by time-varying channels, vehicle mobility, and computing loads. A digital twin (DT) can provide a synchronized virtual representation of the physical vehicular network and update it using real-time system information \cite{gu2025digital}. This enables DT assistance to track system states, predict resource availability, and support proactive optimization before perception performance degrades.

Building on this observation, this paper studies the trade-off between perception accuracy and latency in edge-assisted cooperative perception for autonomous vehicles and develops a DT-assisted optimization framework. To the best of our knowledge, prior work has not jointly considered DT-assisted perception accuracy maximization and end‑to‑end (E2E) latency control under practical communication and resource constraints yet. Different from \cite{hu2022where2comm, urbaningv2x2025}, which study communication-aware feature sharing and perception performance, this work jointly optimizes the communication confidence threshold and transmit power under explicit E2E latency constraints. In the proposed framework, DT-assisted state estimation enables latency-aware adaptation of cooperative perception decisions at roadside unit (RSU) edge servers. Specifically, the DT layer is leveraged to (i) maintain
time-synchronized virtual replicas of vehicles and RSUs, (ii) exploit
historical and real-time data to anticipate short-term physical-network
states that feed the optimization, and (iii) execute the proposed
latency-aware algorithm at the RSU edge server while reducing
synchronization-induced communication overhead.
The main contributions of this paper are summarized as follows:

\smallskip \noindent $\bullet$ We develop a DT-assisted cooperative perception
framework in which the DT layer maintains time-synchronized virtual
replicas of vehicles and RSUs, anticipates short-term physical-network
states, and enables adaptive control of communication confidence
thresholds and transmit power under coupled communication, computation,
and perception constraints.

     \smallskip \noindent $\bullet$ We propose a coarse-to-fine search (CTFS) algorithm to solve the formulated non-convex optimization problem of maximizing perception accuracy under latency and resource constraints.
    
     \smallskip \noindent $\bullet$ We evaluate the proposed approach and demonstrate the impact of communication resource limitations on perception accuracy under latency constraints.

\section{System Model}
%======================Fig==================
\begin{figure*}[!tb]
    \centering
    \includegraphics[scale = 0.5]{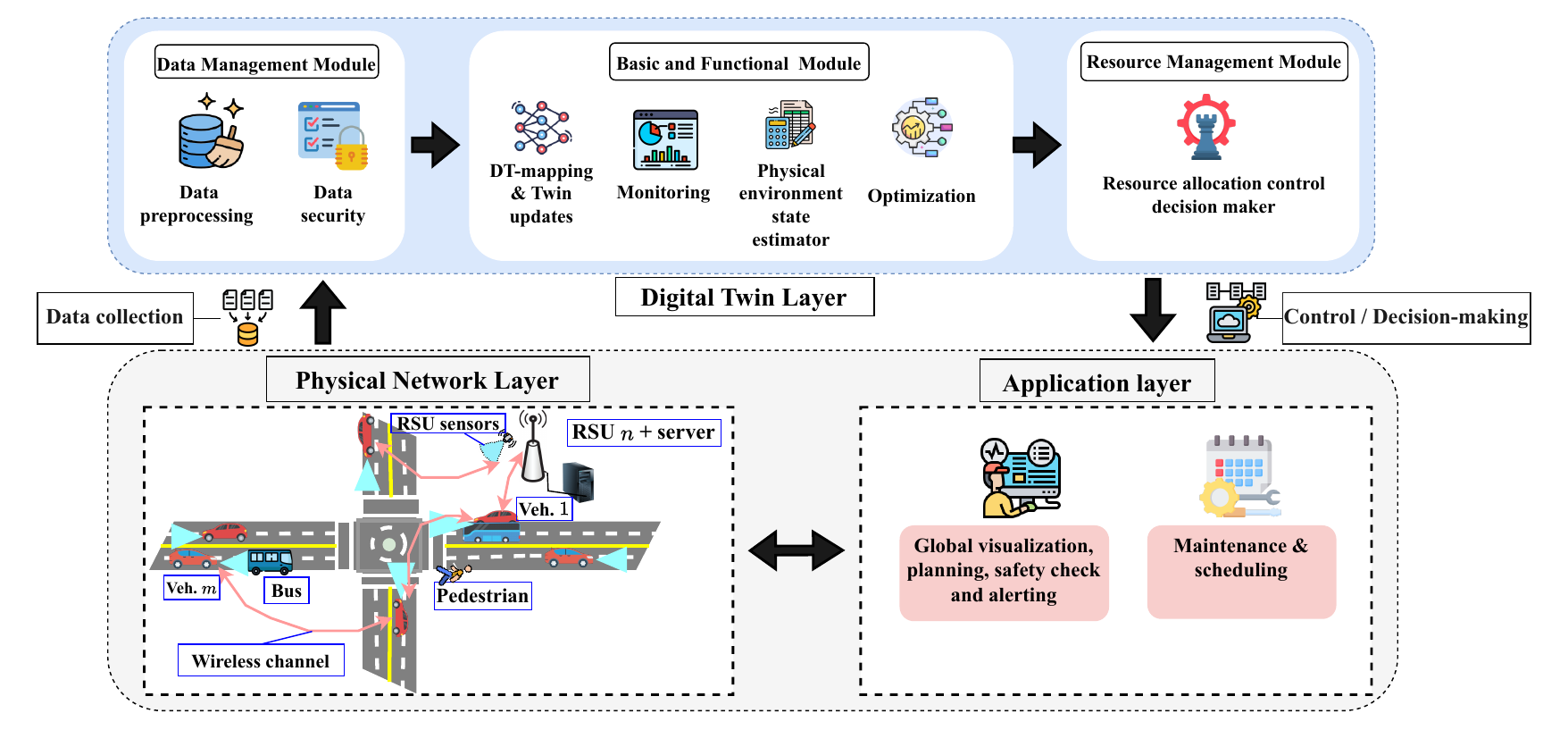}
    \caption{DT framework for cooperative perception in autonomous vehicles.}
    \label{fi222n}
\end{figure*}

\subsection{Physical Network and Feature Extraction Models}

The physical network layer, depicted in Fig. \ref{fi222n}, comprises $M$ autonomous vehicles, represented by the set $\mathcal{M} = \{1, 2,\ldots, M\}$, and $N$ RSUs, denoted as $\mathcal{N} = \{1, 2,\ldots, N\}$. 
Each vehicle is equipped with heterogeneous perception sensors, including cameras and LiDAR sensors, along with onboard computing resources to preprocess raw sensory data and extract relevant features. We focus on autonomous vehicles supported by the vehicle-to-everything (V2X) communication system. These vehicles use sub-6 GHz radio interfaces to transmit their extracted feature data to their associated RSUs via V2X communication. The uplink (UL) transmission from the vehicles to the RSUs is influenced by channel conditions, allocated bandwidth, and transmission power. Each RSU is equipped with an edge server capable of hosting the DT platform and executing latency-sensitive DT functions, including feature aggregation, multi-source data fusion, state synchronization, prediction, and decision support.

\subsubsection{Feature Extraction and Fusion Models}
Following \cite{urbaningv2x2025}, we consider the trained models and datasets used for cooperative perception. The considered framework adopts a multi-modal PointPillars-based feature extraction model implemented through OpenCOOD \cite{OpenCOOD}, where LiDAR point clouds are converted into pseudo-images. It supports early, intermediate, and late fusion strategies using models such as Where2comm \cite{hu2022where2comm} and V2X-ViT \cite{xu2022v2x}. In this work, we focus on intermediate fusion based on the Where2comm model and the PointPillars architecture \cite{urbaningv2x2025, hu2022where2comm, lang2019pointpillars}.

\subsection{DT Layer} \label{DTlayer}
The proposed DT-assisted cooperative perception framework is shown in Fig. \ref{fi222n}. 
The DT layer consists of the following three main modules:
\subsubsection{Data Management Module}

This module collects and processes real-time data from the physical network, including vehicle locations, RSU locations, resource availability, and perception task information. It supports data quality management and secure information exchange, thereby enabling reliable DT updates and real-time network monitoring.

\subsubsection{Basic and Functional Module}

This module constructs and maintains the virtual representation of the physical network. It supports network monitoring, anomaly detection, and DT calibration to improve the consistency between the physical and virtual networks. It also estimates the future states of the physical environment and executes optimization algorithms for determining communication confidence thresholds, transmit power levels, and resource allocation decisions.
\subsubsection{Resource Management Module}
This module uses the outputs of the basic and functional module to determine control decisions and communicate them to the physical network. These decisions include resource allocation and transmission parameter updates that improve system responsiveness under latency constraints.

The above three DT layer modules are deployed within the DT platform hosted at the edge server of RSU \( n \).
Through the basic and functional modules of the DT layer, each physical network element has a corresponding DT model:
 The DT model of vehicle $m$ at time slot $t$ is defined as
\begin{math}
DT_m(t) = \{pos_m(t), DC_m(t), F_m, p_m^{\text{Max}}\},
\end{math}
where $pos_m(t) = \{d_m(t), \upsilon_m(t), a_m(t), \xi_m(t)\}$ includes the vehicle position $d_m(t)$, velocity $\upsilon_m(t)$, acceleration $a_m(t)$, and heading $\xi_m(t)$. The term $DC_m(t) = \{D_m(t), T_m^{\text{Max}}(t)\}$ represents the perception task information, where $D_m(t)$ is the task size and $T_m^{\text{Max}}(t)$ is the maximum tolerable processing time. The parameters $F_m$ and $p_m^{\text{Max}}$ denote the computing capability and maximum transmit power of vehicle $m$, respectively.

The DT model of the edge server at RSU $n$ is given by
\begin{math}
    DT_n = \{pos_n, F_n^{\text{Max}}\},
\end{math}
where $pos_n$ denotes the fixed location of RSU $n$, and $F_n^{\text{Max}}$ denotes its computational capacity.

The application layer in Fig. \ref{fi222n} serves as an intermediary between the DT layer and the physical network layer, managing tasks like global network visualization, safety checks, and maintenance scheduling or alerts.

Initially, each physical network element transmits its real-time operational data to the edge server of its associated RSU, where the corresponding DT model is instantiated, as described above. For newly observed vehicles, the DT model is created at the serving RSU using the latest physical-network data. However, if a vehicle already has an existing DT model and transitions from one RSU to another, we assume that its DT model is smoothly migrated to the newly associated RSU through high-speed wired connections between RSUs, thereby ensuring seamless operation and service continuity during handover, and reducing the cost of building a new DT model from scratch.

The DT layer continuously interacts with the physical network to maintain an updated virtual representation of the physical entities. This interaction is regulated by the synchronization parameter $\pi$, which controls the frequency of data updates between the physical network and the DT layer. Specifically, every $\pi$ time slots, the DT layer requests the most recent real-time data from the physical network and updates the corresponding DT models accordingly, consequently reducing the overhead communication cost. 
Through this periodic synchronization process, the DT layer maintains live, persistent, and time-synchronized virtual replicas of the physical network.

It is worth noting that maintaining the DT layer introduces additional
overhead, including periodic synchronization traffic, computational load
at the RSU edge server for running the estimator and the CTFS algorithm,
and storage cost for persistent DT states. A detailed quantitative
characterization of this overhead is beyond the scope of this work and is
left for future investigation.

\subsection{Deployment and Computing Model}

\subsubsection{Vehicle Feature Extraction Delay Model}
The feature extraction delay of vehicle $m$ at time slot $t$ is modeled based on the feature extraction architectures in \cite{hu2022where2comm, urbaningv2x2025} as
\begin{math}
   \tau_m^{\text{Feat}}(t) = \frac{C_{m,\text{comp}}}{F_m},
\end{math}
where \( C_{m,\text{comp}} \) denotes the total number of floating-point operations (FLOPs) required by the feature extraction model described in \cite{urbaningv2x2025}.  \( C_{m,\text{comp}} \) is determined by the number of layers across all stages of the bird's-eye-view (BEV) network, as well as the input feature grid's height, width, number of filters, and kernel size.
$F_m$ is the computing capability of vehicle $m$.

\subsubsection{Uplink Channel and Delay Model} \label{Sect22}

The wireless channel between vehicle $m$ and RSU $n$ is modeled as
\begin{math}
g_{m,n}(t) = \chi_{m,n}(t)d_{m,n}^{-\gamma}(t),
\end{math}
where $\gamma$ is the path loss exponent, and $d_{m,n}(t)$ is the distance between vehicle $m$ and RSU $n$. The term $\chi_{m,n}(t)\sim \exp(1)$ denotes the small-scale fading power gain, assumed to be exponentially distributed with a mean of 1, i.e., Rayleigh fading channel and it remains constant within a slot and varies independently across different time slots.

The basic and functional module of the DT layer estimates the channel gain and the corresponding achievable data rate over the sub-6 GHz link. The estimated UL data rate from vehicle $m$ to RSU $n$ is given by
\begin{equation} 
\begin{split}
   \mathsmaller{R_{m,n}(t) = B_m \log_2 \left(1+ \frac{p_m(t) g_{m,n}(t)}{N_0}\right),}
\end{split}
\end{equation}
where $B_m$ and $p_m(t)$ denote the bandwidth and transmit power allocated to vehicle $m$, respectively, and $N_0$ is the noise power.

The UL transmission delay is calculated as \cite{li20256g}
\begin{equation}
    \mathsmaller{\tau_m^{\text{Up}}(t) = \frac{(1-\theta_m(t))D_m(t)}{R_{m,n}(t)},}
\end{equation}
where $\theta_m(t) \in [0,1]$ is the communication confidence threshold defined in \cite{hu2022where2comm}. This threshold controls the portion of the feature map selected for transmission to the RSU. 
\begin{math}
D_m(t),
\end{math}
is the feature map size, which is determined by the original input dimensions, and the bit depth of each feature value \cite{urbaningv2x2025}.

\subsubsection{Waiting Delay Model at the RSU}

Before performing synchronization and fusion, the RSU must receive the BEV feature maps from all participating vehicles. We assume that all vehicles in $\mathcal{M}$ participate in the fusion process at RSU $n$. Let $\tau_n^{\text{Start}}(t)$ denote the time at which synchronization and fusion can start at RSU $n$. Since the RSU must wait for the slowest participating agent, this time is expressed as
\begin{equation}
   \mathsmaller{\tau_n^{\text{Start}}(t) = \max \left\{ \tau_1^{\text{Syn}}(t), \tau_2^{\text{Syn}}(t), \cdots, \tau_M^{\text{Syn}}(t), \tau_n^{\text{Feat}}(t) \right\},}
\end{equation}
where $\tau_m^{\text{Syn}}(t) = \tau_m^{\text{Feat}}(t) + \tau_m^{\text{Up}}(t)$ for all $m \in \mathcal{M}$ represents the total feature extraction and UL transmission time of vehicle $m$. $\tau_n^{\text{Feat}}(t)$ denotes the feature extraction time at RSU $n$.

\subsubsection{Computing and Fusion Delay at the RSU}
After receiving the feature maps from all participating vehicles, RSU $n$ performs feature fusion. The fusion delay is modeled as
\begin{equation}
\mathsmaller{\tau_n^{\text{Fuse}}(t)= \frac{C_{n,\text{comp}}}{F_n},}
\end{equation}
where $C_{n,\text{comp}}$ is the total number of FLOPs required by the fusion model of the RSU $n$ detailed in \cite{urbaningv2x2025}, and it is determined by 
backbone processing over the feature maps received from all participating agents,  attention fusion and deblocking/upsampling computations.
$F_n$ is the computing capability of the edge server.

\subsubsection{Downlink Delay}

After fusion at RSU $n$, the detection results are transmitted back to the vehicles for implementation. Although prior studies such as \cite{shi2025joint} often treat this delay as negligible due to the small size of detection outputs, we explicitly account for it by setting $\tau_m^{\text{Down}}(t)=0.01$ seconds as a conservative air-interface/scheduling overhead to avoid assuming instantaneous feedback.

\section{Problem Formulation}

The total end-to-end (E2E) latency experienced by vehicle $m$ accounts for all computation and communication delays involved in cooperative perception and is given by
\begin{equation} \label{effectivedelay}
   \mathsmaller{ \tau_m^{\text{E2E}}(t) = \tau_n^{\text{Start}}(t) + \tau_n^{\text{Fuse}}(t) + \tau_m^{\text{Down}}(t). }
\end{equation}

Let $f_{\theta}: X(\theta_m,t) \longrightarrow \text{mAP}(\theta_m,t)$ denote the trained cooperative perception model used during deployment under a given communication confidence threshold. The input $X(\theta_m,t)$ consists of the feature maps transmitted by all agents at time slot $t$, while the output is the detection performance measured by the per-slot mean average precision (mAP), $\text{mAP}(\theta_m,t)$. The function $f_{\theta}$ is obtained through offline training and is treated as a data-driven mapping during optimization with $\theta = \{1, 2,\ldots, \theta_M\}$.

The objective is to maximize the perception accuracy, $  \text{mAP}(\theta,t) = f_{\theta}(x)$, where $x \in X(\theta_m,t)$,  while penalizing transmission power consumption, and it is captured by the following objective function
\begin{equation} \label{objet}
     \mathsmaller{ obj(\theta,t) =\text{mAP}(\theta_m,t) - \rho \sum_{m=1}^{M} \frac{p_m(t)}{p_m^{\text{Max}}}. }
\end{equation}
Here, \eqref{objet} reflects the trade-off between the detection performance achieved at the RSU through feature fusion and the power expenditure of the vehicles. $\rho$ is the weighting factor,  accounting for the cost associated with transmission power.

The resulting optimization problem is formulated as
\begin{subequations}
\begin{align}
\mathcal{P}\textbf{1}:\quad 
&\max_{\theta_m,\,p_m}\; obj(\theta,t) \\
\text{s.t.}\quad
&\mathsmaller{\tau^{\text{E2E}}_{m}(t) \le T^{\text{Max}}_{m}(t), \forall m \in \mathcal{M},}
\label{C_1}\\
&\mathsmaller{0 < \theta_m(t) \le 1, \forall m \in \mathcal{M},}
\label{C_2}\\
&\mathsmaller{ 0 \le p_m(t) \le p_m^{\text{Max}}, \forall m \in \mathcal{M}.}
\label{C_3}
\end{align}
\end{subequations}

Problem $\mathcal{P}\textbf{1}$ is subject to several constraints. Constraint \eqref{C_1} ensures that the E2E latency for each vehicle does not exceed its maximum tolerable latency. Constraint \eqref{C_2} restricts the communication confidence threshold to a valid range, while constraint \eqref{C_3} enforces the transmit power limitations imposed by vehicle hardware. Due to the nonlinear dependence of the E2E latency and objective function on $\theta_m(t)$ and $p_m(t)$, $\mathcal{P}\textbf{1}$ constitutes a nonlinear and non-convex optimization problem.

\vspace{-2mm}
\section{Proposed Solution}

In problem $\mathcal{P}\textbf{1}$, the metric $\text{mAP}(\theta,t)$ is obtained from a trained cooperative perception model. Since this model does not provide an explicit analytical form with respect to the communication confidence threshold, the first- and second-order derivatives of the objective function are not directly available. Therefore, derivative-based optimization methods, such as gradient descent, are not suitable for solving $\mathcal{P}\textbf{1}$. To address this challenge, we propose a CTFS algorithm that exploits the structure of the problem to identify promising regions of the search space while avoiding unnecessary evaluations. The proposed algorithm consists of the following two stages:

\paragraph*{\textbf{Stage 1: Coarse Search}}

The coarse search stage constructs an initial search space for the communication confidence threshold. Specifically, for each vehicle $m \in \mathcal{M}$, the threshold $\theta_m \in [0,1]$ is discretized into $q$ equally spaced values. The resulting candidate set $\mathcal{Q}$ therefore contains $q^M$ threshold vectors. For example, when $q=6$ and $M=2$, the candidate set contains $36$ threshold vectors and can be represented as
\begin{math}
\mathsmaller{\mathcal{Q} = \{[0.0,0.0], [\cdots], [0.4,0.6], [\cdots], [1.0,1.0]\}.}
\end{math}
Each vector contains the threshold values of the participating vehicles. As $q$ or $M$ increases, the number of candidates grows rapidly, which increases the computational burden of exhaustive evaluation.

For each candidate vector in $\mathcal{Q}$, the minimum transmit power required to satisfy the latency constraint in \eqref{C_1} is computed. By substituting the UL delay expression into the latency constraint, the required power of vehicle $m$ becomes
\begin{equation} \label{optipowerl}
\mathsmaller{
    {p_m^*(t) = \frac{N_0}{g_{m,n}(t)}
    \bigg(2^{\frac{(1-\theta_m(t))D_m(t)}{B_m \tau_{m,\text{Budget}}^{\text{Up}}(t)}} - 1 \bigg),}}
\end{equation}
where $\tau_{m,\text{Budget}}^{\text{Up}}(t)$ is the maximum allowable UL transmission time for vehicle $m$, defined as
\begin{math}
\tau_{m,\text{Budget}}^{\text{Up}}(t)
= T_m^{\text{Max}}(t) - \tau_m^{\text{Feat}}(t)
- \tau_n^{\text{Fuse}}(t) - \tau_m^{\text{Down}}(t).
\end{math} 
This per-vehicle budget provides a tractable feasibility condition for satisfying the E2E latency constraint \eqref{C_1}. The derivation of \eqref{optipowerl} is provided in the Appendix~\ref{appendix}.
If the computed $p_m^*$ violates the transmit power constraint in \eqref{C_3}, the corresponding threshold vector is discarded. Otherwise, the perception accuracy $\text{mAP}(\theta,t)$ is obtained, and the objective value in \eqref{objet} is evaluated and stored. This feasibility check avoids unnecessary perception model evaluations for infeasible candidates and reduces the computational cost of the search.

At the end of Stage 1, the $K$ threshold vectors with the largest objective values are selected. Retaining multiple candidates improves robustness to small variations in the learned perception model and prevents the fine-tuning stage from being restricted to a single potentially suboptimal region.

\paragraph*{\textbf{Stage 2: Fine-Tuning}}

%%%%%%%%%%%%%%%%%%%%%%%%%%
% CTFS algorithm
%%%%%%%%%%%%%%%%%%%%%%%%%%
\IncMargin{1em}
\begin{algorithm}[!tb]
\scriptsize
\caption{Proposed CTFS Algorithm}\label{Algo1}
\SetKwData{Left}{left}\SetKwData{This}{this}\SetKwData{Up}{up}
\SetKwFunction{Union}{Union}\SetKwFunction{FindCompress}{FindCompress}
\SetKwInOut{Input}{Input}\SetKwInOut{Output}{Output}
\Input{Number of vehicles $M$ and RSUs $N$.}
\Output{Optimal communication confidence threshold $\theta^*$ and transmit power $p^*$.}
\BlankLine
\underline{\textbf{Stage 1: Coarse search}}\\
Initialize the set $\mathcal{Q}$ by discretizing $\theta_m \in [0,1]$ into $q$ equally spaced values for each $m \in \mathcal{M}$, resulting in $q^M$ candidate vectors in $\mathcal{Q}$.\\

\For{each candidate vector $\mathcal{Q}(j)$ in $\mathcal{Q}$}{
    Compute the minimum required power $p_m^*$ that satisfies the latency constraint in \eqref{C_1} for the threshold vector $\mathcal{Q}(j)$, as given in \eqref{optipowerl}.\\
    \textcolor{gray}{\# Check the transmit power constraint in \eqref{C_3}}\\
    \If{$p_m^*$ is not feasible}{
        Remove $\mathcal{Q}(j)$ from $\mathcal{Q}$.
    }
    \Else{
        Obtain $\text{mAP}(\mathcal{Q}(j))$.\\
        Evaluate and store the objective value $obj(\theta,t)$ in \eqref{objet}.
    }
}
Select the $K$ candidate vectors with the largest objective values.\\

\underline{\textbf{Stage 2: Fine-tuning}}\\
Reinitialize $\mathcal{Q}$ by discretizing $\theta_m$ within the refined search range determined by the selected $K$ candidate vectors.\\
Repeat the feasibility check and objective evaluation steps used in Stage 1.\\
Find the candidate vector in $\mathcal{Q}$ that achieves the maximum objective value.\\
Return $\theta^*$ and $p^*$.
\end{algorithm}
\DecMargin{1em}
%======================

The fine-tuning stage follows the same feasibility checking and objective evaluation procedure as the coarse search stage, but it operates over a refined search region. This region is determined by the $K$ candidate vectors selected in Stage 1. The threshold values are then discretized again within this smaller interval, and the discretization resolution can be increased by using a larger value of $q$. This refinement improves the precision of the search while keeping the number of objective evaluations manageable.

The output of this stage is the threshold and power pair, denoted by $\theta_m^*$ and $p_m^*$, that maximizes the objective function in \eqref{objet} while satisfying all constraints. A summary of the CTFS procedure is provided in Algorithm \ref{Algo1}. The search complexity is dominated by the number of evaluated threshold vectors and scales as $\mathcal{O}(q^M)$ per stage. Since the proposed method evaluates the perception objective only for feasible candidates and refines the search around the best coarse solutions, it reduces unnecessary computations compared with exhaustive search over a uniformly fine grid.

The CTFS algorithm is executed by the basic and functional module of the DT layer. The resulting optimal parameters are then delivered to the physical network through the resource management module, allowing the vehicles and RSU to implement the selected communication confidence thresholds and transmit power levels.

The DT technology assists the cooperative perception for autonomous vehicles by running the CTFS algorithm and maintaining and tracking object identities, trajectories, uncertainty estimates, map context, infrastructure states, and estimation.
The DT, specifically the physical environment estimator inside the basic and functional module,
exploits its built-in memory and ability to reuse historical and real-time
data to estimate the future state of the physical environment, including
vehicle positions, velocities, and trajectories. This historical state
sequence can support short-term forecasting through trajectory
extrapolation, motion-pattern matching, or predictive analytics. These
estimated quantities, together with the small-scale fading statistics,
allow the DT layer to estimate the future channel gain $g_{m,n}(t)$ and
UL data rate defined in Section~\ref{Sect22}, which are then used as
inputs to the CTFS algorithm for proactive communication confidence threshold and power decisions.
We note that the future states of the physical environment can be predicted using a learning-based prediction model, instead of the estimation approach used in the paper. The design of learning-based predictors is left for future work.

\vspace{-2mm}

\section{Performance Evaluation}

\subsection{Simulation Setup}\label{setup}

We consider a cooperative perception system with two collaborating vehicles, i.e., $M=2$, and one RSU equipped with an edge server, i.e., $N=1$ \cite{urbaningv2x2025}. Each vehicle has a computing capability of $F_m = 2$ TFLOPs/s, while the RSU edge server has a computing capability of $F_n = 25$ TFLOPs/s \cite{lu2025joint}. The communication bandwidth allocated to each vehicle is set to $B_m = 15$ MHz. The maximum transmit power is $p_m^{\text{Max}} = 23$ dBm, and the noise power is $N_0 = -100$ dBm. The path loss exponent is set to $\gamma = 2.0$, to assume the free-space path-loss, yielding a best-case link benchmark.

The maximum allowable E2E latency for each collaborating vehicle is set to $0.1$ s and $0.3$ s to evaluate the effect of different latency requirements. The transmission cost weight is set to $\rho = 1/M$, to assign equal importance to the power cost of each vehicle. Each time slot lasts \(\Delta t = 0.1\) s, corresponding to the frame length/time step in the UrbanIng dataset \cite{urbaningv2x2025}. Details regarding the feature extraction and fusion model parameters can be found in \cite{urbaningv2x2025}.

\subsection{Datasets}\label{datasett}

We consider the cooperative perception dataset and trained models presented in \cite{urbaningv2x2025}. In particular, we focus on the LiDAR 3D point cloud data, following the analysis in \cite{urbaningv2x2025}, where the LiDAR detection range is set from 0 to 100 meters. The cooperative perception model builds on the framework in \cite{hu2022where2comm, urbaningv2x2025}, where the perception model is trained offline before deployment.
During deployment, the size of the transmitted BEV feature map is controlled by optimizing the communication confidence threshold $\theta_m$ for each vehicle $m \in \mathcal{M}$. Unlike \cite{hu2022where2comm, urbaningv2x2025}, which consider bidirectional feature map exchange between vehicles and infrastructure, this work focuses on edge-assisted cooperative perception. In the considered framework, vehicles transmit their extracted features to RSUs equipped with edge servers. The DT layer at the RSU then performs feature fusion and return the detection results to the vehicles.

\subsection{Results and Discussion}\label{RST}

The evaluation reproduces the perception model in \cite{hu2022where2comm} and applies the proposed optimization framework. 
Detection performance is measured using mAP computed over the first 50 time slots, and the standard deviation, which quantifies the variation in accuracy from the mean at each time slot. 
All results are evaluated at an intersection-over-union (IoU) threshold of $0.3$, consistent with \cite{urbaningv2x2025}.

\begin{figure}[!tb] 
    \centering
    \includegraphics[scale = 0.25]{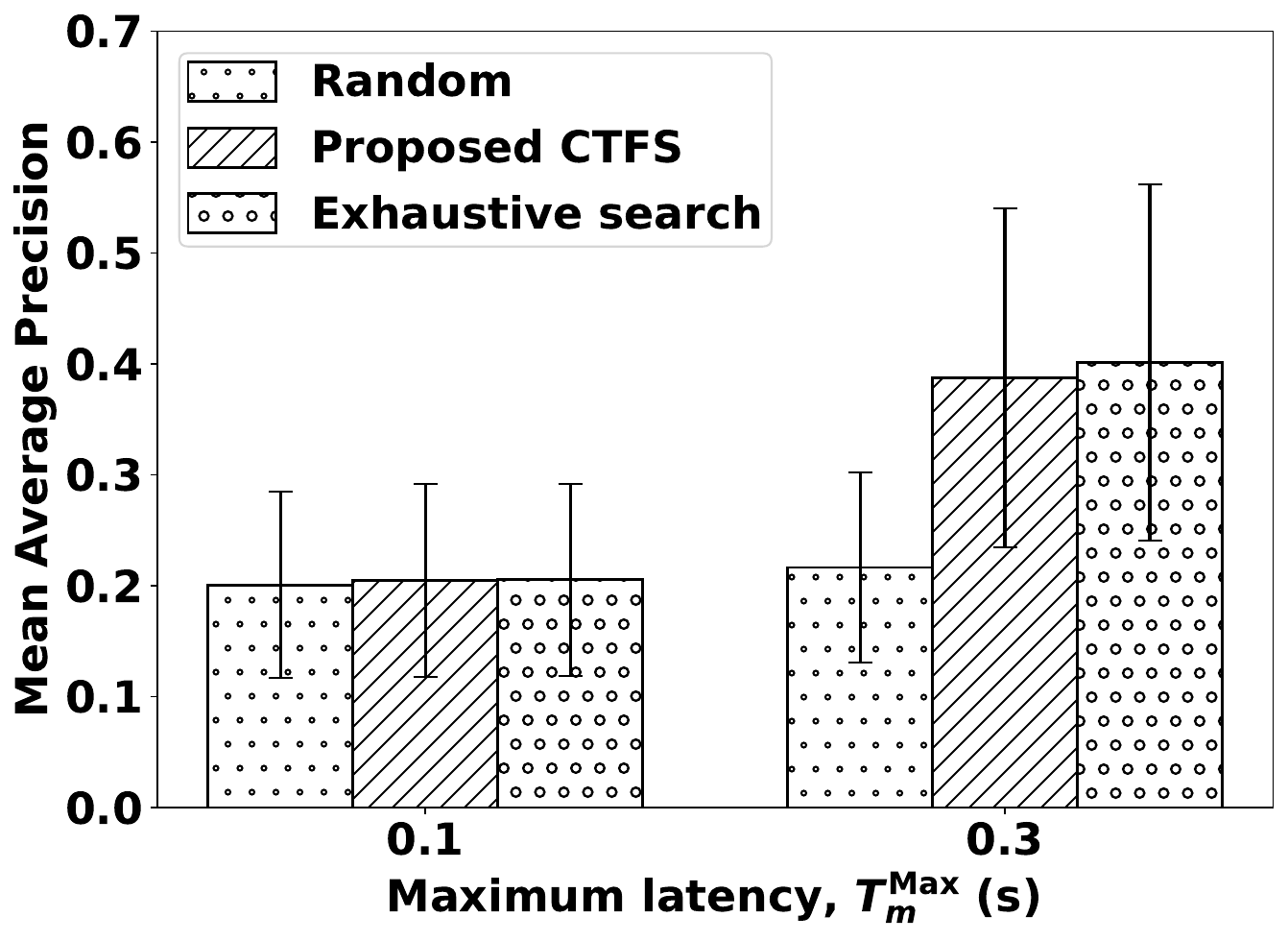}
    \caption{Mean average precision performance under different latency requirements with $B_m = 15$ MHz.}
    \label{fig:gg1}
\end{figure}

Fig. \ref{fig:gg1} compares the mAP achieved by the proposed CTFS algorithm under different latency requirements with exhaustive search and random selection. As the latency constraint becomes more stringent, the achieved mAP decreases across all methods. This behavior arises because tighter latency budgets require the CTFS algorithm to select a larger communication threshold $\theta_m$, which increases feature compression and reduces the quality of BEV information transmitted to the RSU. When $T_m^{\text{Max}} = 0.3$ s, CTFS achieves a mAP of 96.6 \% closer to that of exhaustive search, as sufficient time is available to transmit higher-quality feature maps. In contrast, random search yields noticeably lower mAP due to its indiscriminate exploration of feasible $\theta_m^*$ and $p_m^*$ values. When $T_m^{\text{Max}} = 0.1$ s, the mAP drops significantly, and the performance of all algorithms converges, since the stringent latency constraint prevents the transmission of rich feature representations. These results confirm the inherent trade-off between perception accuracy and latency under limited communication resources.

\begin{figure}[!tb]
    \centering
    \includegraphics[scale = 0.25]{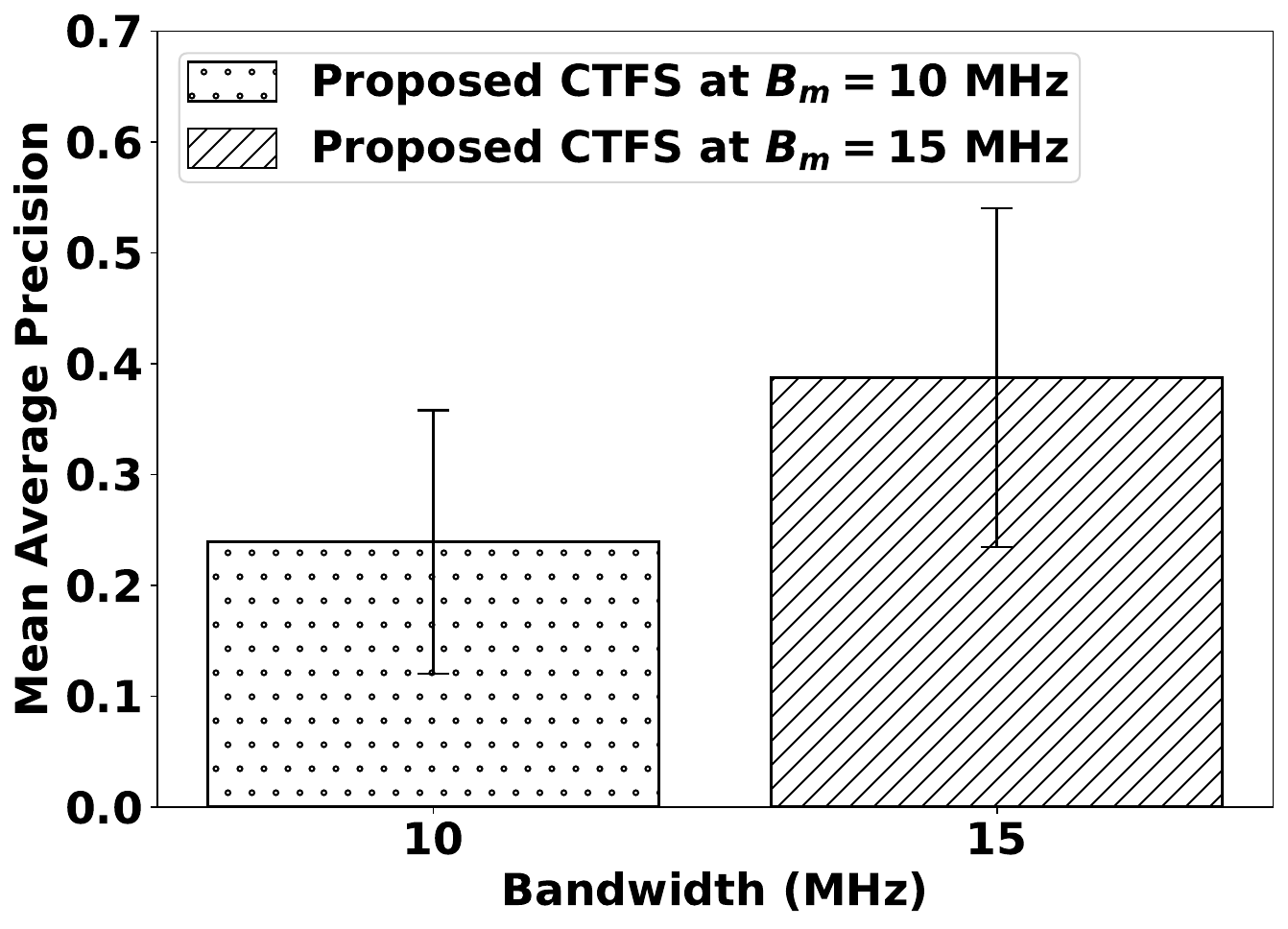}
    \caption{Mean average precision under different bandwidth values with $T_m^{\text{Max}} = 0.3$ s.}
    \label{fig:gg2}
\end{figure}

Fig. \ref{fig:gg2} illustrates the impact of bandwidth allocation on CTFS performance for $T_m^{\text{Max}} = 0.3$ s. Limited bandwidth restricts the volume of BEV feature information that can be transmitted within the latency budget, necessitating stronger compression and leading to reduced mAP after fusion. Conversely, increasing the available bandwidth enables the transmission of richer feature representations and improves the achievable mAP, highlighting the critical role of communication resources in latency-constrained cooperative perception.

The proposed CTFS algorithm also achieves a substantial reduction in computational cost compared with exhaustive search. Its complexity is given by
$\mathcal{O}\left(q^M + 2q^M \left[M C_{m,\mathrm{comp}} + C_{n,\mathrm{comp}}\right]\right)$,
where the factor $2$ accounts for the coarse search and fine-tuning stages. Exhaustive search operates on a uniformly fine grid with complexity
$\mathcal{O}\left(q^M + q^M \left[M C_{m,\mathrm{comp}} + C_{n,\mathrm{comp}}\right]\right)$,
and requires a large $q$ to approximate the continuous optimum. In the considered setup, using $q=30$ for exhaustive search and $q=8$ for CTFS results in $94.95 \times 10^{12}$ and $13.50 \times 10^{12}$ computations, respectively, corresponding to a reduction of approximately $85.78\%$ while maintaining near-exhaustive search performance. Although random search has lower nominal complexity, it consistently exhibits inferior detection performance. Overall, these results highlight the performance–complexity trade-off among the considered schemes, with CTFS offering the most favorable balance by achieving near-exhaustive search performance at a substantially reduced computational cost.

In non-DT cooperative perception, vehicles transmit extracted feature maps and the metadata of the network elements at every time slot, resulting in a communication load of $R_{\text{nonDT}} = \frac{D_m}{\Delta t}$. In contrast, the proposed DT-assisted framework performs estimated, time-synchronized state updates every $\pi$ slots, which reduces the communication load to $R_{\text{DT}} = \frac{D_m}{\pi \Delta t}$. This corresponds to a communication load saving of $1-\frac{1}{\pi}$.
For instance, with $\pi = 2$ time slots, the DT-assisted framework achieves
a $50\%$ reduction in the transmitted data and communication overhead.
The optimization of $\pi$ under mobility- and channel-coherence-aware
criteria is left for future work.
By leveraging persistent state information instead of repeatedly transmitting potentially noisy feature maps and metadata, the DT reduces redundant data exchange and mitigates asynchrony-induced reprocessing.

\section{Conclusion}

This study investigated the trade-off between cooperative perception accuracy and E2E latency in latency-sensitive autonomous driving applications. The results show that cooperative perception performance strongly depends on the availability of communication and computing resources, since accurate perception requires the timely transmission and fusion of high-quality feature information. However, limited resources and stringent latency requirements restrict the amount of perceptual information that can be exchanged, thereby limiting the achievable perception accuracy.

\appendices

\section{Proof of \eqref{optipowerl}}
\label{appendix}

From \eqref{C_1} and \eqref{effectivedelay}, define the UL delay budget of vehicle $m$ as 
$\tau_{m,\text{Budget}}^{\text{Up}}(t)=T_m^{\text{Max}}(t)-\tau_m^{\text{Feat}}(t)-\tau_n^{\text{Fuse}}(t)-\tau_m^{\text{Down}}(t)$. 
For $\tau_{m,\text{Budget}}^{\text{Up}}(t)>0$, the UL delay model in Section \ref{Sect22} must satisfy
\begin{equation}
\mathsmaller{
\frac{(1-\theta_m(t))D_m(t)}
{B_m \log_2 \left(1+\frac{p_m(t)g_{m,n}(t)}{N_0}\right)}
\le
\tau_{m,\text{Budget}}^{\text{Up}}(t).
}
\label{eq_appendix}
\end{equation}
Solving \eqref{eq_appendix} for $p_m(t)$ yields \eqref{optipowerl}, which completes the proof.

%%%% References

\begingroup
\tiny
%\balance
\bibliographystyle{IEEEtran}
\bibliography{References}

@IEEEtranBSTCTL{IEEEexample:BSTcontrol,
	CTLuse_url               = "no",
	CTLdash_repeated_names   = "no",
	CTLuse_forced_etal       = "yes",
	CTLmax_names_forced_etal = "1",
	CTLnames_show_etal       = "1"
}

@inproceedings{li20256g,
  title={{6G THz-semantic access for federated learning enabled internet of unmanned vehicle agents}},
  author={Li, Zihong and Hao, Guozhi and Pan, Zhenni and Wu, Jun},
  booktitle={2025 IEEE 101st Veh. Technol. Conf. (VTC2025-Spring)},
  pages={1--5},
  month = {Jun.},
  year={2025},
  organization={IEEE}
}

@article{urbaningv2x2025,
  title={{UrbanIng-V2X: A Large-Scale Multi-Vehicle, Multi-Infrastructure Dataset Across Multiple Intersections for Cooperative Perception}},
  author={Karthikeyan Chandra Sekaran and Markus Geisler and Dominik Rößle and Adithya Mohan and Daniel Cremers and Wolfgang Utschick and Michael Botsch and Werner Huber and Torsten Schön},
  journal={arXiv [2510.23478]},
  month = {Oct.},
  year={2025}
}

@article{wei2025cooperative,
  title={{Cooperative perception for automated driving: A survey of algorithms, applications, and future directions}},
  author={Wei, Chuheng and Wu, Guoyuan and Barth, Matthew J},
  journal={Proc. IEEE},
  month = {Oct.},
  year={2025},
  publisher={IEEE}
}

@article{ngo2023cooperative,
  title={{Cooperative perception with V2V communication for autonomous vehicles}},
  author={Ngo, Hieu and Fang, Hua and Wang, Honggang},
  journal={IEEE Trans. Veh. Technol.},
  volume={72},
  number={9},
  pages={11122--11131},
  month = {Apr.},
  year={2023},
  publisher={IEEE}
}

@article{bai2024survey,
  title={{A survey and framework of cooperative perception: From heterogeneous singleton to hierarchical cooperation}},
  author={Bai, Zhengwei and Wu, Guoyuan and Barth, Matthew J and Liu, Yongkang and Sisbot, Emrah Akin and Oguchi, Kentaro and Huang, Zhitong},
  journal={IEEE Trans. Intell. Transp. Syst.},
  volume={25},
  number={11},
  pages={15191--15209},
  month = {Sep.},
  year={2024},
  publisher={IEEE}
}

@inproceedings{xu2022v2x,
  title={{V2X-ViT: Vehicle-to-everything cooperative perception with vision transformer}},
  author={Xu, Runsheng and Xiang, Hao and Tu, Zhengzhong and Xia, Xin and Yang, Ming-Hsuan and Ma, Jiaqi},
  booktitle={arXiv [2203.10638]},
  month = {Oct.},
  year={2022}
}

@article{hu2022where2comm,
  title={{Where2comm: Communication-efficient collaborative perception via spatial confidence maps}},
  author={Hu, Yue and Fang, Shaoheng and Lei, Zixing and Zhong, Yiqi and Chen, Siheng},
  journal={arXiv [2209.12836]},
  month = {Dec.},
  year={2022}
}

@article{lu2025joint,
  title={{Joint optimization of compression, transmission and computation for cooperative perception aided intelligent vehicular networks}},
  author={Lu, Binbin and Huang, Xumin and Wu, Yuan and Qian, Liping and Zhou, Sheng and Niyato, Dusit},
  journal={IEEE Trans. Veh. Technol.},
  volume={74},
  number={5},
  pages={8201--8214},
  month = {Jan.},
  year={2025},
  publisher={IEEE}
}

@article{gu2025digital,
  title={{Digital twin technology for intelligent vehicles and transportation systems: A survey on applications, challenges and future directions}},
  author={Gu, Xiaohui and Duan, Wei and Zhang, Guoan and Hou, Jia and Peng, Limei and Wen, Miaowen and Gao, Feifei and Chen, Min and Ho, Pin-Han},
  journal={IEEE Commun. Surv. Tutor.},
  month = {June},
  year={2025},
  publisher={IEEE}
}

@inproceedings{lang2019pointpillars,
  title={{Pointpillars: Fast encoders for object detection from point clouds}},
  author={Lang, Alex H and Vora, Sourabh and Caesar, Holger and Zhou, Lubing and Yang, Jiong and Beijbom, Oscar},
  booktitle={IEEE Conf. Comput. Vis. Pattern Recognit.},
  pages={12697--12705},
  year={2019}
}

@misc{OpenCOOD,
  author  = {OpenCOOD},
  title   = {{Welcome to OpenCOOD’s documentation!, Readthedocs.io. [Online]. Available:}},
  howpublished ={\url{https://opencood.readthedocs.io/en/latest/}},
  note = {[Accessed: 29-Apr-2026]}
}

@article{shi2025joint,
  title={{Joint task offloading and channel allocation in spatial-temporal dynamic for MEC networks}},
  author={Shi, Tianyi and Zhang, Tiankui and Loo, Jonathan and Huang, Rong and Wang, Yapeng},
  journal={IEEE Trans. Industr. Inform.},
  volume = {21},
  pages = {6240–6250},
  month = {May},
  year={2025},
  publisher={IEEE}
}
\endgroup

\end{document}